\documentclass[11pt]{article}

\usepackage{epsfig,amssymb,amstext}

\newtheorem{platz}{{\bf Fig.}} 

\newcommand{\FIGo}[3]{%
\marginpar{ \begin{platz} \label{#1} ~ \end{platz} \vspace*{1.5ex} }}
\newcommand{\printfigcap}[2]{{\bf Figure~\ref{#1}}: #2
\goodbreak\vspace{0.5ex}}

\begin{document}


\thispagestyle{plain}
\noindent
{\Large A  Computational Procedure to Detect a New Type of}\\[2ex] 
{\Large High Dimensional Chaotic Saddle and its 
Application to }\\[2ex] 
{\Large the 3-D Hill's  Problem} \\[5ex]
\begin{center}
{\large H Waalkens, A Burbanks and  S Wiggins}\\[4
ex]
{\large School of Mathematics\\
        Bristol University\\ University Walk\\ 
        Bristol, BS8 1TW , UK}\\[2ex]

\today
\end{center}

\vspace*{1ex}

\section*{Abstract}

A  computational procedure that allows the detection of a new type of high-dimensional chaotic saddle in Hamiltonian systems with three degrees of freedom is presented.
The chaotic saddle is associated with a so-called {\em normally hyperbolic invariant manifold} (NHIM). 
The procedure allows to compute appropriate homoclinic orbits to the NHIM from which we can infer  the existence a
chaotic saddle.
NHIMs control the phase space transport across an equilibrium point of saddle-centre-...-centre stability type,
which is a fundamental mechanism for chemical reactions, capture and escape, scattering,  and, more generally,  ``transformation'' in many different areas of physics.
Consequently, the presented methods and results are of broad interest.
The procedure is illustrated for the spatial Hill's problem which
is a well known model in celestial mechanics and which
gained much interest e.g. in the study of the formation of
binaries in the Kuiper belt.

\vspace{3ex}

\noindent PACS: 05.45.Jn, 45.50.Pk\\


\section*{Introduction}

Chaotic saddles are the saddle-type invariant Cantor sets associated with a
horseshoe construction \cite{nusse}. They play a central role in many complex dynamical phenomena, e.g., the existence of supertransients \cite{lai1} and the fractal structure of chaotic scattering \cite{chaos}.
The chaotic saddles constructed to date have
been related to either homoclinic orbits associated with hyperbolic
periodic orbits or special types of equilibria. In this letter we are concerned
with a fundamentally new  type of high dimensional chaotic saddle structure that can occur only in Hamiltonian systems with three or more
degrees of freedom (DOF), and we describe a computational method for detecting it which also illustrates its geometrical complexity. 

The chaotic saddle is associated with a {\em normally hyperbolic invariant manifold} (NHIM) \cite{Wiggins3}.
The physical significance of NHIMs arises from the fact that their stable and unstable manifolds control the 
{\em phase space transport} across
equilibria of saddle-centre-...-centre stability type. This is
 is not only the fundamental mechanism for the evolution from reactants to products in chemical reaction,   
but also for ``transformations'' in general in a large, and diverse, number of applications as e.g.   
ionisation problems in atomic physics \cite{JaffeFarellyUzer2000}, rearrangements of
clusters \cite{KomaBerry1999}, cosmology \cite{cosmo}, and solid state and semi-conductor physics \cite{Jacucci,Eckhardt1995}.
Though it had been recognised that it is important to understand the {\em dynamics}
near saddle-centre-...-centre equilibria   
it was only recently that new developments in dynamical systems theory offered the theoretical framework 
and computing power offered the means to study the {\em phase space structure} near saddle-centre-...-centres for 
systems with 3 or more DOF \cite{Wiggins1,Wiggins2,wwju,ujpyw}.
Besides the NHIM and its stable and unstable manifolds it is now possible to compute a codimension 1 submanifold of the energy surface, the so-called {\em transition state}, which is transverse to the Hamiltonian flow and {\em locally} divides the energy surface into two disjoint components. 
{\em Locally}, the transition state provides the only means of passing from one phase space region (associated with ``reactants'') to another phase space region (associated with ``products''), i.e. trajectories must cross the transition state in order to ``react''.
The transversality of the transition state to the Hamiltonian flow 
is essential for rate calculations as it solves the problem of {\em locally} recrossing trajectories. 

In this letter we go beyond this local result and consider more global issues associated with the dynamics related to the transition state and NHIM.
In particular, we describe a computational method for determining  the existence of homoclinic and heteroclinic connections to NHIMs. Using recent results  in \cite{Cress1},  the existence of certain types of these homoclinic orbits allows us to infer the existence of a new type of chaotic saddle (in fact, a Cantor set of chaotic saddles). Because of the ubiquity of saddle-centre-...-centre type equilibria in applications (as described above), we expect that the 
methods and results presented here, which are applicable to three and more DOF, to be of broad interest.

The physical system we choose to illustrate our method is one
from celestial mechanics; the 3-d Hill's equations \cite{Hill}.
Advances in detector technology have opened up new frontiers in
celestial mechanics with the discovery of trans-Neptunian objects and
binary systems in the Kuiper belt. 
The calculation of capture probabilities
requires the study of the 3-d Hill's equations (rather than the thoroughly
studied 2-d case) because many of the capture events occur from high
inclination (see e.g. the recent work by Goldreich, Lithwick and Sari \cite{Gold}).


\section*{Hill's Problem and the Phase Space Structure Near  Saddle-centre-centre Equilibria}

The circular restricted three body problem (CRTBP) models the motion
of a tiny particle under the gravitational influence of one (large)
primary mass and one (smaller) secondary mass both in circular orbits
about their common centre of mass \cite{murray_dermott}.  Hill's problem is a limit version
of the CRTBP which describes the motion of the particle in a neighbourhood of the secondary
mass.  Hill's equations can be derived from the following Hamiltonian
in dimensionless units,
\[ 
H = \frac{1}{2} ( p_x^2 + p_y^2 +p_z^2 ) + 
y p_x - x p_y 
-x^2
+\frac{1}{2} ( y^2 + z^2 )
-\frac{3}{r}  
\]
where $r=(x^2+y^2+z^2)^{1/2}$. 
It is well known that Hill's equations have two equilibria, which are denoted
traditionally by L$_1$ and L$_2$, and that the matrices associated with linearising Hamilton's equations about each equilibrium have a pair of real eigenvalues and two pairs of pure imaginary complex conjugate eigenvalues. This means
that L$_1$ and L$_2$ are equilibria of saddle-centre-centre type.

A detailed theory for {\em phase space transport} near
saddle-centre-centre equilibrium points has been developed in recent
years \cite{Wiggins1,Wiggins2,wwju,ujpyw}. For energies slightly above that
of the saddle-centre-centre equilibrium point, on the $5$-dimensional
energy surface there exists an invariant $3$-dimensional sphere $S^3$
of saddle stability type.  This $3$-sphere is significant for two
reasons:

\begin{itemize}
\item
It is the ``equator'' of a $4$-dimensional sphere called the {\em
transition state}.  Except for the equator (which is an invariant
manifold), the transition state is locally a ``surface of no return''
in the sense that all trajectories that start in a neighbourhood of the
transition state in the energy surface must cross the transition state
and exit the neighbourhood (in the appropriate direction in time).  For
energies ``sufficiently close'' to the energy of the
saddle-centre-centre equilibrium point, the transition state satisfies
the {\em bottleneck property}.  This means that, 
the energy surface {\em locally} has the geometrical structure of $S^{4} \times
I$ (i.e., 4-sphere $\times$ interval) and the transition state divides the
energy surface into two disjoint components.  Moreover, the {\em only}
way a trajectory can pass from one component of the energy surface to
the other is to pass through the transition state.

\item
The 3-sphere is a {\em normally hyperbolic invariant manifold} \cite{Wiggins3} (NHIM),
which means that the expansion and contraction rates of the dynamics
on the 3-sphere dominate those transverse to it.  Just like a ``saddle
point'' it therefore has stable and unstable manifolds. In this case
the stable and unstable manifolds are $4$-dimensional, having the
structure of {\em spherical cylinders}, $S^{3} \times \mathbb R$.
Hence, they are of one less dimension than the energy surface and act
as ``separatrices''; they ``enclose'' a volume of the energy
surface. Their key dynamical significance is that the only way that
trajectories can pass through the transition state is if they are
inside the region of the energy surface enclosed by the stable and
unstable spherical cylinders.

\end{itemize}

These phase space structures can be realised through a procedure based
on Poincar{\'e}-Birkhoff normalisation by which explicit formulae for
the NHIM, its stable an unstable manifolds, and the transition state
are given \cite{ujpyw} in ``normal form coordinates''.  The phase
space structures are then mapped back into the original coordinates by
the inverse of the normal form (NF) transformation.

\section*{Computation and Visualisation of the
Phase Space Structures near L$_1$ and L$_2$}

We will visualise the phase space structures near Hill's equilibria
L$_1$ and L$_2$ as their projections to configuration space.  In
understanding the result, and its implications, it is useful to recall
that the level sets of the effective potential energy
\[ 
V \equiv H - \frac12 |(p_x+y,p_y-x,p_z)|^2 = 
-\frac32x^2 + \frac12 z^2 - \frac3r,
\]
the so-called {\em zero velocity surfaces}, confine the motion in
configuration space.

In figure~\ref{fig:zvs}(a) we show the zero velocity surface (ZVS) $V=E$
for an energy $E=-4.4$ which is 0.1 above the energy of the
equilibrium points. The ZVS encloses the energetically-allowed volume
of configuration space. For Hill's equations its shape has two
``bottlenecks'' associated with L$_1$ and L$_2$ which divide the
allowed volume into three regions: a bounded region about the origin
(for which $-1<x<+1$ (A)) and two unbounded regions (for which $x<-1$ (B) and
$x>+1$ (C), respectively).  The phase space structures near L$_1$ and
L$_2$ regulate the transport through the bottlenecks between these
regions.

\def\figzvs{%
(a) Cut-away of the zero velocity surface (blue/green).
(b) Transition state at L$_2$ (red) and NHIM at L$_1$ (white).
The energy is -4.4\,.
}
\def\FIGzvs{
\centerline{
\raisebox{5cm}{(a)}
\includegraphics[angle=0,width=5cm]{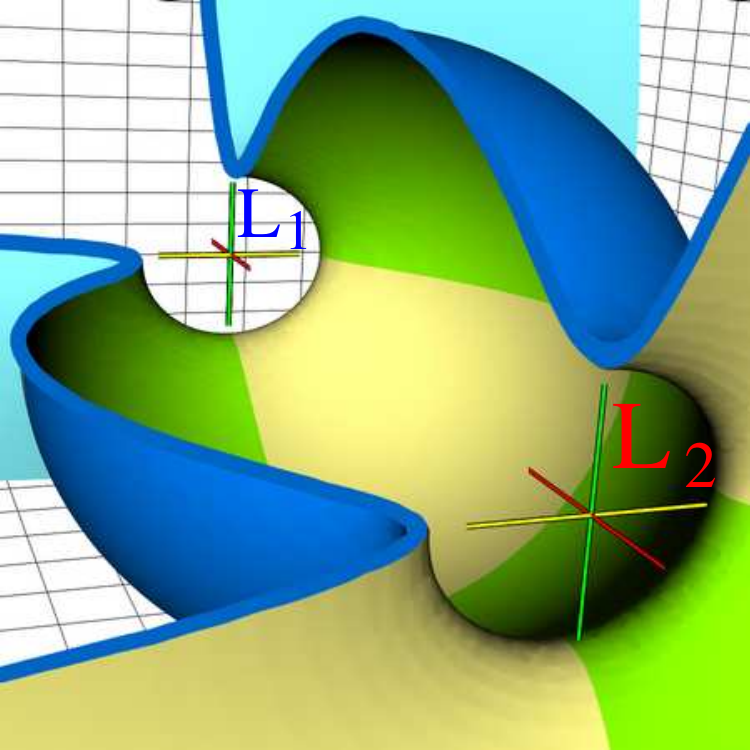}
\hspace*{1cm}
\raisebox{5cm}{(b)}
\includegraphics[angle=0,width=5cm]{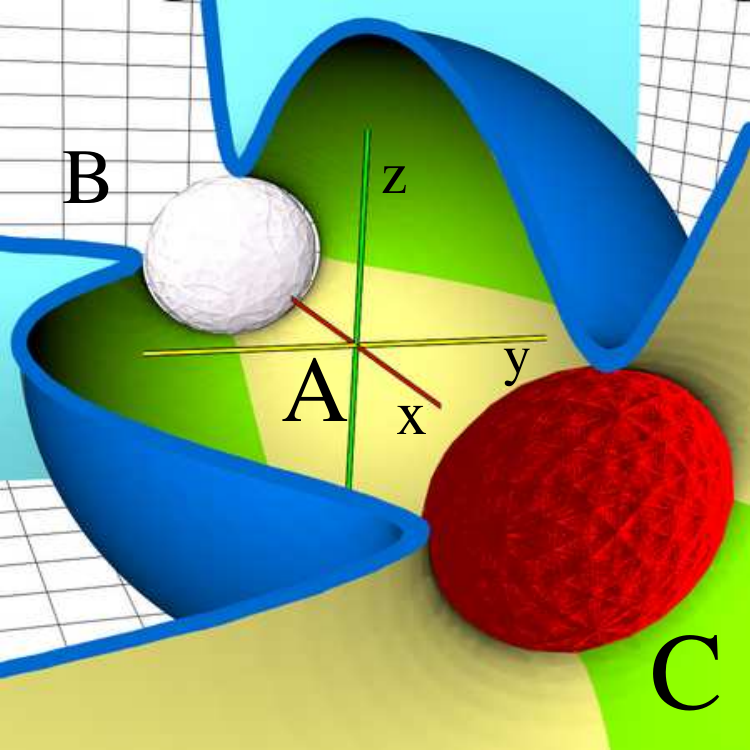}
}
}
\FIGo{fig:zvs}{\figzvs}{\FIGzvs}


The NFs about L$_1$ and L$_2$ are related by symmetry. It is thus
sufficient to compute explicitly only the NF about L$_1$.  As a result of 
the two complex eigenvalues associated with the equilibria being rationally independent, the NF, to
any desired finite order of computation, is completely integrable with
integrals given by ${\cal I} = p_1 q_1$, $J_i = \frac{1}{2} \left( q_i
^2 + p_i^2 \right), \, i=2, 3$. The NF Hamiltonian can be written solely
as a function of the integrals, $H = H({\cal I}, J_2, J_3)$, and
Hamilton's equations decouple into the product of
independent linear systems
\[
(\dot{q}_1,\dot{p}_1)=  \frac{\partial H}{\partial {\cal I}} (q_1,-p_1),\,
(\dot{q}_i,\dot{p}_i)=  \frac{\partial H}{\partial J_i} (p_i,-q_i),\,
i=2,3\,.
\]
The transition state 4-sphere is given by $q_1=p_1$. From the
equations of motion 
it is easy to see that it is a ``surface of no return''.  Its equator
$p_1=q_1=0$ is the NHIM ($3$-sphere). It has $4$-dimensional stable
($q_1=0$) and unstable ($p_1=0)$ manifolds.
In the energy surface volume enclosed by the stable and unstable manifolds ${\cal I}$ is positive; outside ${\cal I}$ is negative.
The NHIM has a special structure. It is {\em foliated} by a one
parameter family of invariant $2$-tori (the ``Hopf fibration'')
\cite{Wiggins1,Wiggins2,ujpyw}. These tori can be parametrised e.g. by
the integral $J_2$ where $J_3$ is then given implicitly by energy
conservation and ${\cal I}=0$. At the minimal and maximal values of
$J_2$ the 2-tori degenerate to periodic orbits, the so-called ``Lyapunov orbits''.  
The $2$-tori and the periodic orbits have
$3$-dimensional and $2$-dimensional stable and unstable manifolds,
respectively, which are contained in the $4$-dimensional stable
and unstable manifolds of the NHIM. These geometrical considerations
will play an important role described below.

We perform the NF computation to order 18 using the computer algebra
system {\em Mathematica}.  As a result the NF Hamiltonian is a sum
over $219$ multivariate monomials in $({\cal I},J_2,J_3)$. Each
component of the mapping between NF coordinates and the original phase
space coordinates involves sums over about $50\,000$ multivariate
monomials. The resulting transition state and NHIM are shown in
figure~\ref{fig:zvs}(b). Notice that the transition state completely blocks
the ``bottleneck'' in the ZVS in configuration space; it also blocks
the bottleneck in the level set of the Hamiltonian in phase space.

\section*{Computation of Homoclinic and Heteroclinic
Connections between the NHIMs near L$_1$ and L$_2$}

The ``saddle integral'', ${\cal I} = p_1 q_1$, plays the key role in
our numerical approach to detecting orbits connecting the same NHIM,
or orbits connecting different NHIMs.
Taking into account the Hopf fibration of the NHIM our procedure consists of the following four steps:

\begin{itemize}
\item
In the NF coordinates, choose an invariant $2$-torus on the NHIM and
seed a mesh covering the torus with initial conditions. Displace these
initial conditions slightly in the direction of the unstable manifold
of this torus ($p_1=0, \, q_1 = \varepsilon$).

\item
Map the initial conditions back into the original coordinates using
the NF transformation.

\item
Integrate the initial conditions forward in time using Hill's
equations. Since they are in the unstable manifold they will leave the
neighbourhood in which the NF transformation is valid (which is why
we integrated them in the original coordinates).

\item
Check if a trajectory returns to the neighbourhood of L$_1$ or L$_2$
where the NF is valid. If so, map it into the NF coordinates and check
the value of its saddle integral. If the saddle integral is zero, the
trajectory must be on the stable manifold of the respective NHIM
(since the trajectory is already on the unstable manifold and two
unstable manifolds cannot intersect).
\end{itemize}

\def\fighopf{%
The top panels highlight individual 2-tori in the Hopf fibration of the NHIM near L$_1$ and shows
contours of values of the saddle integral ${\cal I}_{\text{L}_1}$ and ${\cal I}_{\text{L}_2}$ on the 2-tori. The 2-tori are for $J_2=n J_{2\,\text{max}}/4$, $n=1,2,3$, where $J_{2\,\text{max}}$ is the
maximum $J_2$ on the NHIM.  $n=0$ and $n=4$ correspond to the two
Lyapunov periodic orbits.  The 2-tori are parametrised by the angles
$\alpha_2$ and $\alpha_3$ conjugate to $J_2$ and
$J_3$.  ($E= -4.4$.) For clarity the bottom panels show the tori in the covering space.}
\def\FIGhopf{
\centerline{
\includegraphics[angle=0,width=15cm]{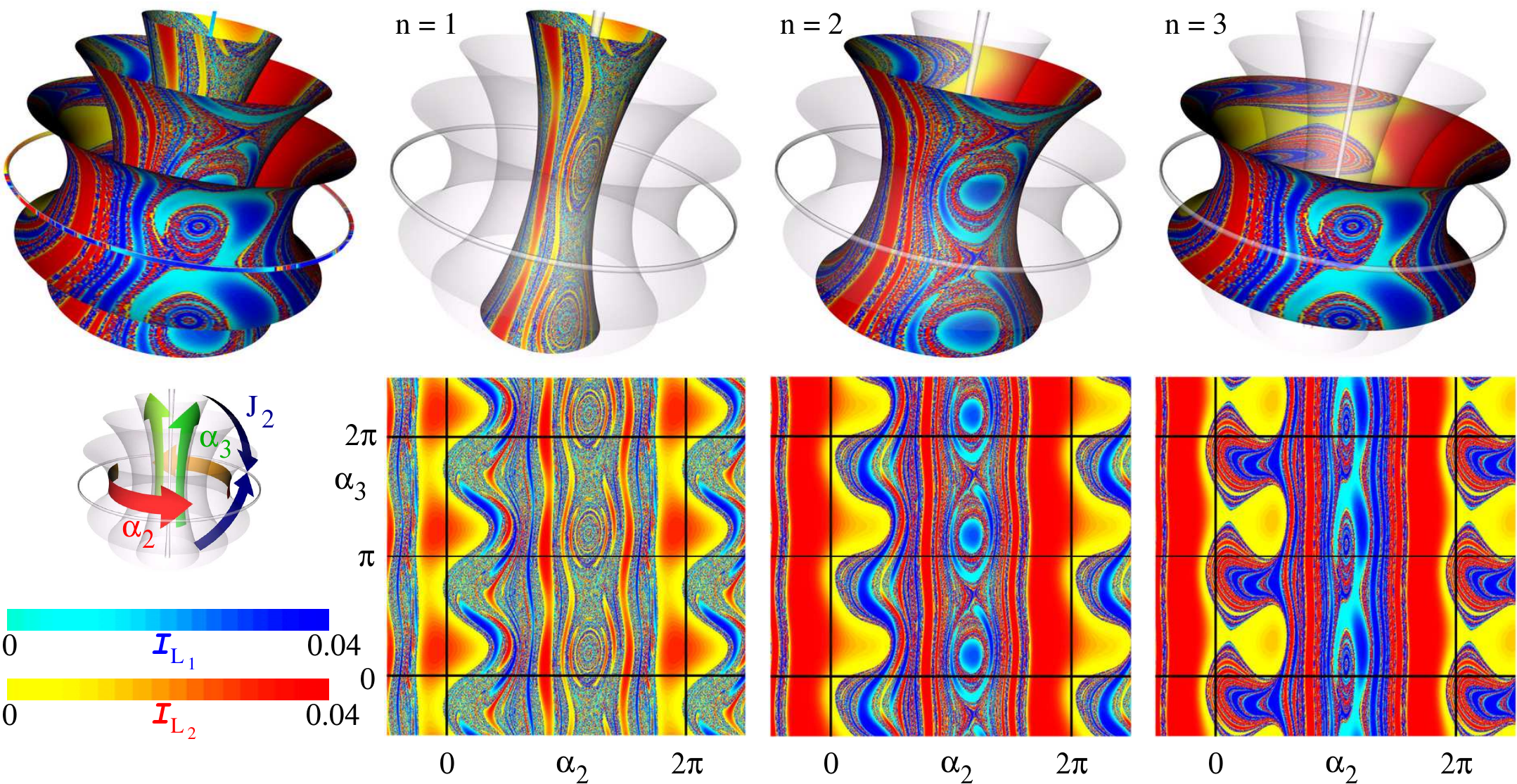}
}
}
\FIGo{fig:hopf}{\fighopf}{\FIGhopf}


The procedure can be understood as a shooting method between the
${\cal I}$-fibres of the locally valid NFs.  We apply the method to
Hill's problem by displacing (with $\epsilon=10^{-4}$) meshes of initial conditions on invariant
2-tori of the NHIM near L$_1$ along the respective unstable manifold
branches which are directed towards the bounded region about the
origin, see figure~\ref{fig:zvs}.  These initial conditions are
propagated by integrating Hill's equations where the singularity at
the origin is taken care of by the Kustaanheimo-Stiefel
regularisation.  If the resulting trajectory reenters or enters the
neighbourhood of validity of the NF about L$_1$ or L$_2$, respectively,
we check the value of the integral ${\cal I}$. A positive ${\cal I}$
means that the trajectory will cross the respective transition state
which leads to an exit to the outside of the bounded region about the
origin.  In this case the integration is stopped. A negative value of
${\cal I}$ means that the trajectory does not exit on this approach of
L$_1$ or L$_2$ and the integration is continued until the trajectory
reaches a validity neighbourhood of the NF with positive ${\cal
I}$. It is to be noted that along the part of a trajectory which
traverses the validity neighbourhood of the NF the values of the
integrals $({\cal I},J_2,J_3)$ are conserved to 12 digits and more,
demonstrating the high accuracy of the NF.

We illustrate the results in figure~\ref{fig:hopf} where contours of the
integral ${\cal I}_{\text{L}_1}$ (light blue/dark blue; exit through
the transition state near L$_1$) and ${\cal I}_{\text{L}_2}$
(yellow/red; exit through the transition state near L$_2$) are shown
plotted on the corresponding torus of the NHIM.  At the boundary of a
light blue/dark blue region ${\cal I}_{\text{L}_1}$ goes to zero and
points on the boundary correspond to homoclinic orbits which connect
the NHIM near L$_1$ to itself. Similarly, boundaries of yellow/red
regions represent heteroclinic connections between the NHIMs near
L$_1$ and L$_2$.

\def\figlyapunov{%
Homoclinic and heteroclinic orbits
between the planar Lyapunov orbit near L$_1$ and the NHIMs near L$_1$ (zeros of blue arcs) and L$_2$ 
(zeros of red arcs).}
\def\FIGlyapunov{
\centerline{
\includegraphics[angle=0,height=5cm]{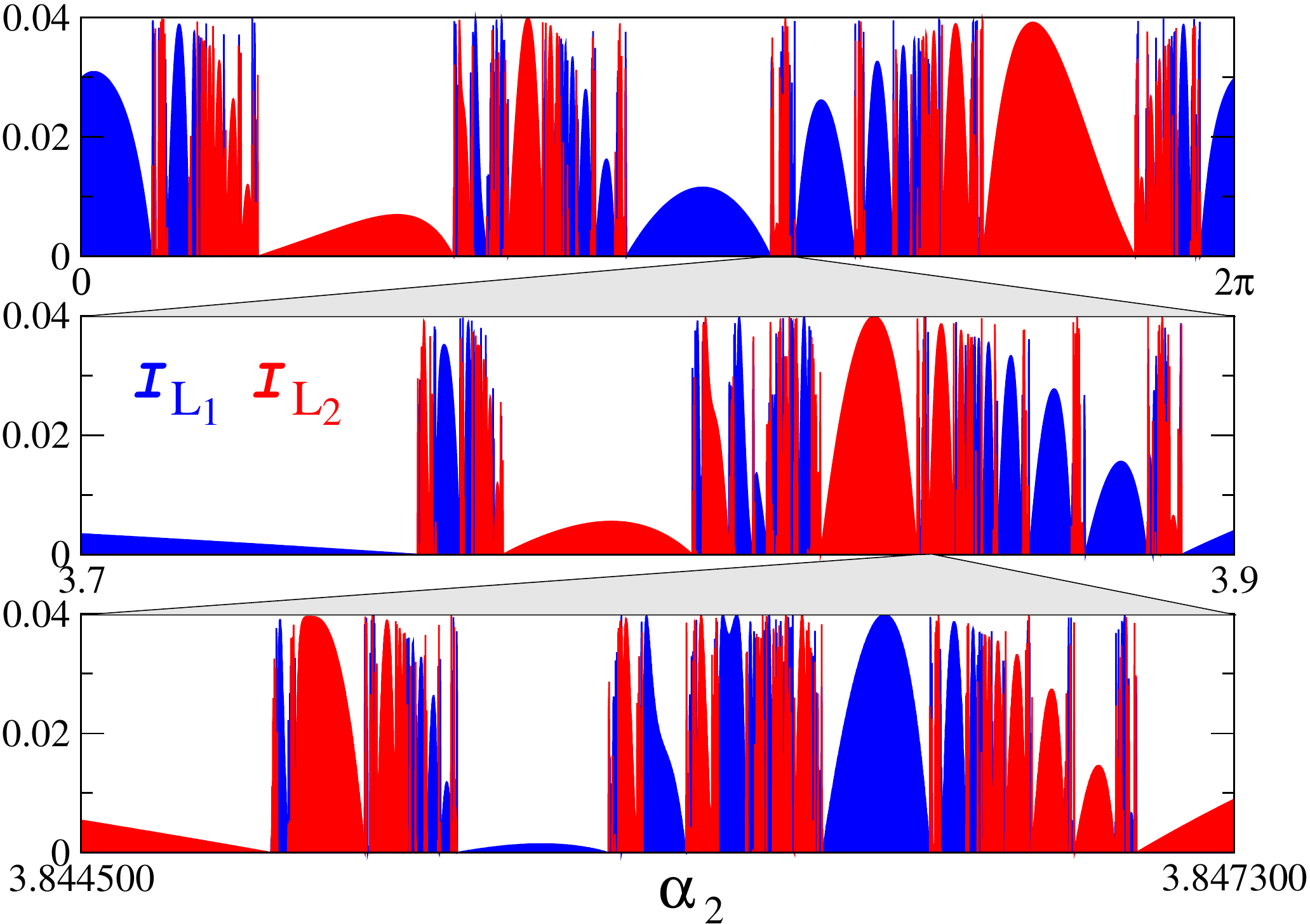}
}
}
\FIGo{fig:lyapunov}{\figlyapunov}{\FIGlyapunov}


Though the individual regions in figure~\ref{fig:hopf} themselves are regular in the sense that
they each have a smooth boundary, the disposition of the regions is
very intricate.  We illustrate this for initial conditions on the
2-dimensional unstable manifold of the Lyapunov periodic orbit near
L$_1$ which has $J_3=0$ and in configuration space lies in the
$(x,y)$-plane, see figure~\ref{fig:lyapunov}. Between each two arcs
which extend from two consecutive zeros of ${\cal I}$ as a function of
the angle conjugate to $J_2$ there is an infinity of further arcs.
Each zero represents an orbit which in time is backward asymptotic to
the Lyapunov periodic orbit and forward asymptotic to the NHIM near
L$_1$ (blue) or the NHIM near L$_2$ (red).  The result is a
self-similar structure well known from classical scattering systems.

\section*{Homoclinic Connections and Chaotic Saddles}

 While orbits homoclinic to normally hyperbolic invariant
tori have been studied (\cite{Wiggins4}), tori {\em cannot} be
normally hyperbolic in {\em Hamiltonian} systems \cite{bolotin}. The NHIM is normally
hyperbolic, yet there are no theorems describing the dynamics
associated with orbits homoclinic to a normally hyperbolic invariant
sphere (part of the difficulty here comes from the fact that a sphere
cannot be described by a single coordinate chart, but see \cite{cosmo} for numerical evidence that this could be an important mechanism for chaos). An important piece
of this problem has recently been solved by Cresson \cite{Cress1} who
proved that if the stable and unstable manifolds of a torus in the
NHIM intersect transversely, then there exists a (uniformly)
hyperbolic invariant Cantor set on which the dynamics is conjugate to
a shift map, i.e., a {\em chaotic saddle}.  The stable and unstable
manifolds of a $2$-torus are $3$-dimensional in the $5$-dimensional
energy surface. A transverse intersection is necessarily
$1$-dimensional, i.e., a trajectory.

\def\fighomorbit{%
(a) Points of homoclinic connections on the 2-torus $J_2=J_{2\,\text{max}}/2$ (see figure~\ref{fig:hopf}). 
(b) A sample homoclinic trajectory in configuration space.}
\def\FIGhomorbit{
\centerline{
\raisebox{5cm}{(a)}
\includegraphics[angle=0,width=5cm]{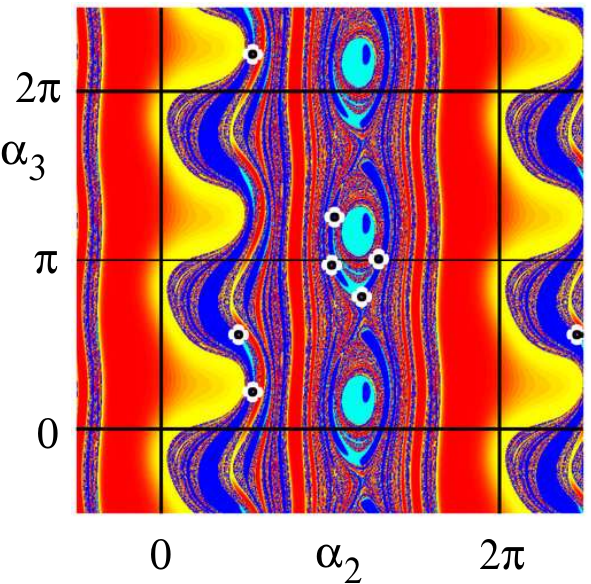}
\hspace*{1cm}
\raisebox{5cm}{(b)}
\includegraphics[angle=0,width=5cm]{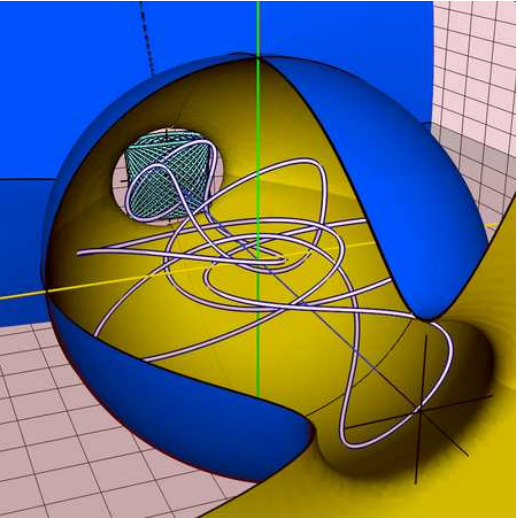}
}
}
\FIGo{fig:homorbit}{\fighomorbit}{\FIGhomorbit}


When the effect of the neglected terms in the normal form expansion (i.e., the non-normalised ``tail'' of the expansion) are included we expect, by KAM theory, that a Cantor set of non-resonant tori in the Hopf fibration will persist.
Homoclinic intersections of the stable and unstable manifolds of a
given 2-torus (rather than more general connections between the
manifolds of such a torus and those of the entire NHIM) can be
detected easily by our numerical procedure described above.  We
therefore not only check the value the integral ${\cal I}$ upon
entering the neighbourhood of validity of the same NF that we started
from, but also the difference between the centre integral $J_2$ of the
2-torus we started from and the centre integral $J'_2$ in the entered
validity neighbourhood.  If both $J_2-J'_2$ and ${\cal I}$ vanish we
have a homoclinic orbit to the 2-torus under consideration.  The
result of such a procedure is shown in figure~\ref{fig:homorbit}(a)
where homoclinic connections appear as the intersection points of the
zero contours of $J_2-J'_2$ (sharp boundaries between dark blue and
light blue within the blue regions) and those of ${\cal I}$
(boundaries of the blue regions
themselves). Figure~\ref{fig:homorbit}(b) shows the example of a
homoclinic orbit which corresponds to one of the prominent
intersection points marked by a white dot in the figure. In fact there
are many more intersection points located in the finer structures of
figure~\ref{fig:homorbit}(a). There is a tendency that
the homoclinic orbits become more complicated in the finer regions.


\section*{Conclusions}
In this letter we have described a numerical method for detecting a Cantor set of chaotic saddles in Hamiltonian systems with
three DOF. We have illustrated it for the spatial Hill's problem.  The implications for this problem should be of current 
interest in celestial mechanics. Recently it has been shown that chaos plays an important role in the energetics of capture 
\cite{nature}, and this Cantor set of chaotic saddles should be central to this process.
Chaotic scattering in systems with three or more DOF is poorly understood, and our results and methods should
provide a window into this subject. Towards this end, our methods detect the chaotic saddles by detecting
appropriate homoclinic orbits. 

Concerning transition state theory, as described by Truhlar \cite{truhlar}, there are two types of trajectories that exhibit
multiple recrossings of the transition state: local and global. As described in the introduction, the correct choice of transition
state solves the local recrossing problem. However, characterising the global recrossing problem, i.e. the question of whether there are trajectories, which, after crossing the transition state and leaving its neighbourhood, return to the transition state 
and cross it again, is more difficult. Under additional assumptions on the geometry of the homoclinic or heteroclinic orbits, the chaotic saddles consist of trajectories that enter and leave a neigborhood of the transition state, recrossing it infinitely often. This will be the subject of a future publication.

It would also be of some interest to actually compute and visualise the chaotic saddles themselves. A recently developed numerical method \cite{sweet} may play a role for this purpose.
Finally, we remark that our results are not limited to 3 DOF. The general theory developed in 
\cite{Wiggins1,Wiggins2,wwju,ujpyw} applies to Hamiltonian systems with arbitrary DOF. The numerical method
generalises easily since the saddle integral is a {\em scalar} function regardless of the number of DOF.

\section*{Acknowledgments}

We thank J~Palacian and P~Yanguas for Mathematica code which we
modified for the NF computation. This work was supported by the Office of Naval Research.
H.W. acknowledges support from the Deutsche Forschungsgemeinschaft (Wa 1590/1-1).


\newpage

\section*{Figure Captions}

\vspace*{1cm}

\printfigcap{fig:zvs}{\figzvs}
\vspace*{0.5cm}
\noindent
\printfigcap{fig:hopf}{\fighopf}
\vspace*{0.5cm}
\noindent
\printfigcap{fig:lyapunov}{\figlyapunov}
\noindent
\printfigcap{fig:homorbit}{\fighomorbit}

\end{document}